\documentclass{agujournal2019}

\usepackage{lineno, layouts, microtype}

\usepackage{bm}
\usepackage{xcolor}
\usepackage{xfrac}
\usepackage{url}
\usepackage[T1]{fontenc} 

\usepackage{amsfonts, amssymb, amsmath}
\usepackage[labelfont=bf]{caption}

\usepackage{comment}

\journalname{Geophysical Research Letters}

\usepackage{scalerel, stackengine}
\setstackEOL{\#}
\stackMath
\def\hatgap{2pt}
\def\subdown{-2pt}
\newcommand\reallywidehat[2][]{ \renewcommand\stackalignment{l} \stackon[\hatgap]{#2}{ \stretchto{
    \scalerel*[\widthof{$#2$}]{\kern-.6pt\bigwedge\kern-.6pt}
    {\rule[-\textheight/2]{1ex}{\textheight}}}
    {0.5ex}_{\smash{ \belowbaseline[\subdown]{\scriptstyle#1} }}
}}

\renewcommand{\arraystretch}{1.5}

\renewcommand{\b}[1]    {\boldsymbol{#1}}
\renewcommand{\r}[1]    {\mathrm{#1}}

\renewcommand{\d}       {\partial}
\newcommand{\bu}        {\b{u}} 

\newcommand{\bxh}       {\hspace{0.1em} \b{\hat x}} 
\newcommand{\byh}       {\hspace{0.1em} \b{\hat y}} 
\newcommand{\bzh}       {\hspace{0.1em} \b{\hat z}}

\newcommand{\half}      {\tfrac{1}{2}}
\newcommand{\beq}       { \begin{linenomath*} \begin{equation}}
\newcommand{\eeq}       {\end{equation} \end{linenomath*}}
\newcommand{\beqs}      {\begin{gather}}
\newcommand{\eeqs}      {\end{gather}}
\newcommand{\balign}    {\begin{linenomath*} \begin{align}}
\newcommand{\ealign}    {\end{align} \end{linenomath*}}

\newcommand{\defn}      {\ensuremath{\stackrel{\r{def}}{=}}}

\newcommand{\bnabla}    {\b{\nabla}}
\newcommand{\nablah}    {\bnabla_{\! \! h}}

\newcommand{\bcdot}     {\b{\cdot}}

\newcommand{\com}       {\, ,}
\newcommand{\per}       {\, .}

\newcommand{\buL}       {\bu^\mathrm{L}}

\newcommand{\uL}        {u^\mathrm{L}}
\newcommand{\vL}        {v^\mathrm{L}}
\newcommand{\wL}        {w^\mathrm{L}}

\newcommand{\buS}       {\bu^\mathrm{S}}
\newcommand{\uS}        {u^\mathrm{S}}
\newcommand{\vS}        {v^\mathrm{S}}

\newcommand{\buE}       {\bu^\mathrm{E}}
\newcommand{\uE}        {u^\mathrm{E}}
\newcommand{\vE}        {v^\mathrm{E}}

\newcommand{\X}         {\mathcal{X}}
\newcommand{\Ro}        {\mathrm{Ro}}

\newcommand{\quarterdegree}{\sfrac{1}{4}^\circ}

\hypersetup{
	breaklinks = true,
	colorlinks = true,
	linkcolor  = Blue,
    urlcolor   = Blue,
	citecolor  = RoyalPurple,
	pdfauthor  = {Gregory L. Wagner, Navid C. Constantinou, and Brandon G. Reichl},
 }

\begin{document}
\justify

\title{
Stokes drift should not be added to ocean general circulation model velocities
}

\authors{
Gregory L. Wagner\affil{1}, Navid C. Constantinou\affil{2}, and Brandon G. Reichl\affil{3}
}

\affiliation{1}{Earth, Atmospheric, and Planetary Sciences, Massachusetts Institute of Technology, Cambridge, MA, USA}
\affiliation{2}{Research School of Earth Sciences \& ARC Centre of Excellence for Climate Extremes,\\[-0.5em] Australian National University, Canberra, ACT, Australia}
\affiliation{3}{NOAA Geophysical Fluid Dynamics Laboratory, Princeton, NJ, USA}

\correspondingauthor{Gregory L. Wagner}{gregory.leclaire.wagner@gmail.com}


\begin{keypoints}
\item Adding Stokes drift to ocean general circulation model velocities is inconsistent with the wave-averaged Craik-Leibovich equations
\item Wave-agnostic general circulation models can simulate total Lagrangian-mean velocities even when Stokes drift is significant
\end{keypoints}

%
%

\begin{abstract}
Studies of ocean surface transport often invoke the ``Eulerian-mean hypothesis': that wave-agnostic general circulation models neglecting explicit surface waves effects simulate the Eulerian-mean ocean velocity time-averaged over surface wave oscillations.
Acceptance of the Eulerian-mean hypothesis motivates reconstructing the total, Lagrangian-mean surface velocity by \textit{adding} Stokes drift to model output.
Here, we show that the Eulerian-mean hypothesis is inconsistent, because wave-agnostic models cannot accurately simulate the Eulerian-mean velocity if Stokes drift is significant compared to the Eulerian-mean or Lagrangian-mean velocity.
We conclude that Stokes drift should not be added to ocean general circulation model velocities.
We additionally show the viability of the alternative ``Lagrangian-mean hypothesis'' using a theoretical argument and by comparing a wave-agnostic global ocean simulation with an explicitly wave-averaged simulation.
We find that our wave-agnostic model accurately simulates the Lagrangian-mean velocity even though the Stokes drift is significant.

\noindent \textbf{Plain language summary} \\
\noindent
Physical oceanographers are taught that surface waves ``induce'' a time-averaged current called the Stokes drift.
This notion motivates studies in which the total ocean surface transport of things like trash, oil, and kelp is estimated by the combined effect of ``ocean currents'' as simulated by an ocean model, or estimated from observations, and an \textit{additional} ``surface wave Stokes drift''.
In this paper, we show that ocean models and observations likely estimate total ocean transport \textit{including} Stokes drift.
So, we shouldn't ``add Stokes drift'' to model output or certain kinds of observations.
\end{abstract}

%
%

\section{Introduction} \label{sec:intro}

Ocean surface waves complicate observations and models of near-surface ocean transport.
Surface waves are associated with significant, yet oscillatory fluid displacements that must be time-averaged away to reveal the underlying persistent circulation.
But time-averaging over surface waves is not straightforward: the ocean velocity averaged at a fixed position --- the ``Eulerian-mean velocity'' --- is missing a component of the total transport called the ``Stokes drift'' \cite{stokes1847theory}.
The total mean velocity responsible for advecting tracers, particles, and water parcels is called the ``Lagrangian-mean velocity'', because it can be obtained by time-averaging currents in a semi-Lagrangian reference frame that follows surface wave oscillations.
These statements are summarized by the timeless formula
\beq \label{total-velocity}
    \buL = \buE + \buS \com
\eeq
where $\buL$ is the surface-wave-averaged Lagrangian-mean velocity, $\buE$ is the surface-wave-averaged Eulerian-mean velocity, and $\buS$ is the surface wave Stokes drift \cite{longuet1969transport}.
On average, tracers, particles, and water parcels follow streamlines traced by Lagrangian-mean velocity $\buL$.
(Formulas analogous to~\eqref{total-velocity} also apply to velocities averaged over longer time intervals, such as supermonthly timescales over mesoscale ocean turbulence, but we do not discuss ``other'' Lagrangian-mean velocities in this paper.)

Most general circulation models of ocean transport and many observation-based estimates based on dynamical balances neither resolve surface wave oscillations nor invoke an explicit dependence on the surface wave state.
Such ``wave-agnostic'' estimates should be \textit{interpreted} as somehow time-averaged over surface wave oscillations.
Note that the expression ``wave-agnostic'' excludes observations based on explicit averaging, such as moored Eulerian velocity measurements, or fully Lagrangian drifter or tracer-based estimates \cite<for in depth discussions and examples see>{longuet1969transport, middleton1989skew, smith2006observed}, that lack the ambiguity inherent to wave-agnosticism.
We ask: do wave-agnostic models estimate the Eulerian-mean velocity, or the Lagrangian-mean velocity?

Studies investigating surface wave effects on ocean transport 
\cite{
    kubota1994mechanism,
    tamura2012stokes,
    fraser2018antarctica,
    iwasaki2017fate,
    van2017stokes,
    dobler2019large,
    onink2019role,
    kerpen2020wave,
    van2020physical,
    bosi2021role,
    van2021dispersion,
    durgadoo2021strategies,
    cunningham2022role, 
    chassignet2021tracking,
    onink2022,
    herman2022typical}
often assume both the Stokes drift makes a significant contribution to total ocean surface transport, \text{and} that wave-agnostic models and observational products estimate the Eulerian-mean velocity.
We call these assumptions the ``Eulerian-mean hypothesis''.
Within the context of the Eulerian-mean hypothesis, the total Lagrangian-mean transport is constructed by adding an estimate of the Stokes drift velocity (derived from an estimate of the surface wave state) to model output or observational products, according to~\eqref{total-velocity}.

In this paper we argue that the Eulerian-mean hypothesis is inconsistent, in that its two components --- accurate Eulerian-mean wave-agnostic simulations, in the presence of significant Stokes drift --- are incompatible.
Our argument, developed in section~\ref{sec:theory}.1, uses a scaling analysis of the explicitly-wave-averaged Eulerian-mean Boussinesq equation \cite{craik1976rational, huang1979surface}.
Because the Eulerian-mean hypothesis is internally inconsistent, ocean transport studies based on wave-agnostic model output or observations based on dynamical balances should not ``add Stokes drift'' to construct the total Lagrangian-mean transport.

To resolve the status of wave-agnostic models, we propose the alternative ``Lagrangian-mean hypothesis'', which supposes that wave-agnostic models estimate the Lagrangian-mean velocity.
The Lagrangian-mean hypothesis is consistent in that it does not contain mutually incompatible components, and is therefore the only viable interpretation of wave-agnostic model output.
The consistency of the Lagrangian-mean hypothesis does not, however, guarantee its \textit{accuracy}.
For example, wave-agnostic and explicitly wave-averaged large eddy simulations at meter scales exhibit dramatic, well-documented differences known as ``Langmuir turbulence'' \cite{skyllingstad1995ocean, mcwilliams1997langmuir, sullivan2010dynamics}.
We therefore turn to the question of whether wave-agnostic general circulation models --- with scales much larger than the scales of large eddy simulation --- provide accurate approximations of Lagrangian-mean general circulation models in section~\ref{sec:theory}.2.
Using a scaling argument, we predict that for \textit{geostrophic} flows, surface wave terms in the Lagrangian-mean wave-averaged equations are negligible.
This suggest that, at mesoscales and larger, the wave-agnostic equations accurately approximate the Lagrangian-mean wave-averaged equations and the Lagrangian-mean hypothesis may hold.

We test the Lagrangian-mean hypothesis in section~\ref{sec:wave-model} by comparing output from a wave-agnostic ``control'' general circulation ocean model simulation with $\sfrac{1}{4}^\circ$ resolution that neglects surface wave effects on velocity and tracers with a ``wave-averaged'' general circulation ocean model simulation that explicitly includes surface waves.
We find that the velocity in the wave-agnostic simulation is almost identical to the Lagrangian-mean velocity in the wave-averaged simulation.
In section~\ref{sec:discussion}, we discuss the implications of our results for surface boundary layer parameterizations and the potential uses of wave-averaged general circulation models.

The success of the Lagrangian-mean hypothesis --- even in a limited context --- cannot be reconciled with the notion that Stokes drift is ``wave-induced transport''.
This underlines the importance of regarding the total Lagrangian-mean momentum as the fundamental momentum variable, rather than the Eulerian-mean momentum.
In such a paradigm, surface waves play no \textit{direct} role in transport, and affect mean dynamics only through wave-averaged forces in the Lagrangian-mean momentum equation that are negligible at large oceanic scales.

\section{Wave-averaged and wave-agnostic dynamics}
\label{sec:theory}

The wave-averaged Craik--Leibovich Boussinesq momentum equation \cite{craik1976rational, huang1979surface}
can be written either in terms of the Eulerian-mean velocity $\buE$,
\beq
\begin{split} \label{uE-momentum}
    \d_t \buE
           + \left ( \buE \bcdot \bnabla \right ) \buE
        &  + f \bzh \times \left ( \buE \; {\color{PineGreen} + \, \buS} \right ) + \\[1ex]
        & \bnabla \left ( \bar p \; {\color{PineGreen} + \, \tfrac1{2} \buS \bcdot \buS + \buS \bcdot \buE } \right )
        = \bar b \bzh
         + \b{\X} 
        {\color{PineGreen} \; + \; \buS \times ( \bnabla \times \buE ) } \com
\end{split}
\eeq
\emph{or} the Lagrangian mean velocity, $\buL$,
\beq \label{uL-momentum}
    \d_t \buL
         + \left ( \buL \bcdot \bnabla \right ) \buL
         + \left ( f \bzh  {\color{PineGreen} \; - \, \bnabla \times \buS } \right ) \times \buL
          + \bnabla \bar p
        = \bar b \bzh
         + \b{\X}
        {\color{PineGreen} \; + \; \d_t \buS } \per
\eeq
In \eqref{uE-momentum}--\eqref{uL-momentum}, $\bar p$ is the Eulerian-mean kinematic pressure (pressure scaled with ocean's reference density), $\bar b \defn - g \rho' / \rho_0$ is the Eulerian-mean buoyancy defined in terms of gravitational acceleration $g$, reference density $\rho_0$, and the Eulerian-mean density perturbation $\rho'$, $f$ is the Coriolis parameter, and $\bzh$ is the unit vector pointing up.
$\b{\X}$ parametrizes subgrid momentum flux divergences associated with, for example, ocean surface boundary layer turbulence.
We discuss $\b{\X}$ further in section~\ref{sec:discussion}.
Equations \eqref{uE-momentum}--\eqref{uL-momentum} are related by~\eqref{total-velocity} and standard vector identities. 
Physical interpretations for the {\color{PineGreen}green} surface wave terms in equations~\eqref{uE-momentum}--\eqref{uL-momentum} are discussed by \citeA{wagner2021near} in their section 2.1, \citeA{buhler2014waves} in their section 11.3.2, and by \citeA{suzuki2016understanding}. 

The {\color{PineGreen}green} surface wave terms in equations~\eqref{uE-momentum} and~\eqref{uL-momentum} depend \emph{explicitly} on the Stokes drift $\buS$ and therefore the surface wave state.
The green terms distinguish equations~\eqref{uE-momentum}--\eqref{uL-momentum} from the wave-agnostic Boussinesq momentum equation,
\beq \label{control-momentum}
    \d_t \bu + \left ( \bu \bcdot \bnabla \right ) \bu + f \bzh \times \bu + \bnabla p = b \bzh + \b{\X} \com
\eeq
solved by typical, wave-agnostic ocean general circulation models.




\subsection{The Eulerian-mean hypothesis is inconsistent}

The Eulerian-mean hypothesis posits that velocities $\bu$ that solve equation~\eqref{control-momentum} are identical or similar to $\buE$ in \eqref{uE-momentum} at ocean mesoscales and larger.
The Eulerian-mean hypothesis therefore requires that \eqref{control-momentum}  is a good approximation to \eqref{uE-momentum} when $\buS \sim \buE$. 

The central flaw in the Eulerian-mean hypothesis is that Stokes-Coriolis term $f \bzh \times \buS$ in~\eqref{uE-momentum} is the same magnitude as the ``Eulerian-mean component of the Coriolis force'', $f \bzh \times \buE$ when $\buS \sim \buE$.
Thus for dynamics close to geostrophic and Ekman balance, \eqref{uE-momentum} is not a good approximation to~\eqref{control-momentum} because it does not represent the \textit{total} Coriolis force $f \bzh \times \buL$.
Note that the Lagrangian-mean nature of the surface-wave-averaged Coriolis force, Ekman balance, and geostrophic balance is known \cite{mcwilliams1999wave, samelson2022wind} and is also feature of internal-wave-averaged quasi-geostrophic dynamics \cite{wagner2015available, kafiabad2021wave}.
Because Coriolis force acts on the Lagrangian-mean velocities, the only consistent interpretation of observations based on Ekman balance, like \citeA{johannessen2016globcurrent}, is that they estimate the Lagrangian-mean velocity.
We finally note that a similar inconsistency in the Eulerian-mean hypothesis applies to tracer advection by $\buL$: $\buE \bcdot \bnabla c$ is not a close approximation to $\buL \bcdot \bnabla c$ when $\buS \sim \buE$.

The failure of the Eulerian-mean hypothesis to account for both tracer advection and the total Coriolis force is sufficient motivation to pursue the Lagrangian-mean hypothesis, and convinced readers may skip to section~\ref{Lagrangian-mean-hypothesis}.
The remainder of this section shows that the ``vortex force'' $\buS \times \left ( \bnabla \times \buE \right )$ and ``Stokes-Bernoulli'' terms aside the pressure in \eqref{uE-momentum} are $O(\Ro)$, where
\beq
    \Ro \defn \frac{U}{f L} \com
\eeq
is the Rossby number for flows with velocity scale $| \buL | \sim | \buS | \sim | \buE | \sim U$ and horizontal scales $L \sim U / | \nablah \bu |$.
We take $U$ and $L$ to apply to both mean currents and Stokes drift.
$\Ro$ is typically less than unity for oceanic motion at mesoscales and larger.

Under slowly-modulated surface waves, the ratio 
\beq \label{wave-modulation}
    \frac{ | \nablah \buS |}{| \d_z \buS |} \sim \frac{H}{L} \com
\eeq
is small, where $H$ is the vertical decay scale of the Stokes drift.
The approximation~\eqref{wave-modulation} simplifies the vortex force in~\eqref{uE-momentum} to
\beq \label{vortex-force}
    \buS \times (\bnabla \times \buE) \approx \vS (\d_x \vE - \d_y \uE) \bxh - \uE (\d_x \vE - \d_y \uE) \byh - \left ( \uS \d_z \uE + \vS \d_z \vE \right ) \bzh \com
\eeq
where $\bxh$ and $\byh$ are unit vectors in horizontal directions.

We simplify the scaling analysis by reusing $H$ and $L$ in \eqref{wave-modulation} for vertical and horizontal near-surface velocity scales.
For the $x$-component of~\eqref{vortex-force} we find
\beq \label{uE-terms-scaling}
    \frac{\d_x \left ( \half \buS \bcdot \buS + \buS \bcdot \buE \right )}{f \vE} \sim
    \frac{\left ( \d_x \vE - \d_y \uE \right ) \vS}{f \vE} \sim \frac{U^2 / L}{f U} = \Ro \per
\eeq
A similar result holds for the $y$-component of~\eqref{vortex-force}.
Compared to the geostrophic pressure gradient $\d_z \bar p \sim f U L / H$, we find that the vertical component of \eqref{vortex-force} scales with
\beq \label{horizontal-stokes-momentum-E}
    \frac{\uS \d_z \uE + \vS \d_z \vE}{\d_z \bar p} \sim \frac{U^2 /H}{f U L / H} = \Ro \per 
\eeq
In summary, in nearly geostrophic mesoscale flows, the Stokes--Coriolis term in \eqref{uE-momentum} is $O(1)$ and non-negligible, which means that \eqref{uE-momentum} is a poor approximation to \eqref{control-momentum} and casts doubt on the Eulerian-mean hypothesis.
The other surface wave terms in \eqref{uE-momentum} are $O(\Ro)$ and are thus negligible for $\Ro \ll 1$.

\subsection{The Lagrangian-mean hypothesis is consistent}
\label{Lagrangian-mean-hypothesis}

The ``Lagrangian-mean hypothesis'' posits that velocities $\bu$ that solve \eqref{control-momentum} are similar to Lagrangian-mean velocities $\buL$ that solve \eqref{uL-momentum} at ocean mesoscale and larger.
We argue that the Lagrangian-mean hypothesis is consistent with a scaling analysis that suggests the green terms in \eqref{uL-momentum} are negligible at ocean mesoscales and larger.

Using~\eqref{wave-modulation} we simplify the surface wave term in~\eqref{uL-momentum},
\beq \label{pseudovorticity}
    (\bnabla \times \buS) \times \buL \approx \wL \d_z \uS \bxh + \wL \d_z \vS \byh - \left ( \uL \d_z \uS + \vL \d_z \vS \right ) \bzh \per
\eeq
The term in~\eqref{pseudovorticity} has the same form as the ``non-traditional'' component of the Coriolis force associated with the horizontal components of planetary vorticity (which have been neglected \textit{a priori} from \eqref{uL-momentum}).
Thus the terms in \eqref{pseudovorticity} are small for the same reason we make the traditional approximation for Coriolis forces: because of the dominance of hydrostatic balance, and because geostrophic vertical velocities scale with
\beq
\wL \sim \Ro \frac{H}{L} U \com
\eeq
and are therefore miniscule at ocean mesoscales and larger where both $\Ro$ and especially $H/L$ are much smaller than unity.
Specifically, the same arguments leading to~\eqref{uE-terms-scaling} conclude that the horizontal components of~\eqref{pseudovorticity} scale with $\Ro^2$ --- much smaller than $O(1)$ and smaller even than the $O(\Ro)$ terms in~\eqref{uE-terms-scaling}.
The vertical component of~\eqref{pseudovorticity} shares the same scaling with~\eqref{horizontal-stokes-momentum-E}: $O(\Ro)$ and therefore negligible at ocean mesoscales and larger.

We save the discussion of $\d_t \buS$ for last.
Only the horizontal components of $\buS$ are significant \cite{vanneste2022stokes}.
$\d_t \buS$ is primarily associated with wave growth beneath atmospheric storms and thus effectively represents the small part of the total parameterized air-sea momentum transfer that is \textit{depth-distributed} rather than fluxed at or just below the surface \cite{wagner2021near}.
We could therefore interpret $\d_t \buS$ as accounted for implicitly in wave-agnostic models by bulk formulae for air-sea momentum transfer.
Even so, we consider a scaling argument by introducing an average $\langle \, \bcdot \,\rangle$ over a time-scale $T$ much longer than a day, and therefore much larger than $f^{-1}$.
We find that
\beq \label{averaged-stokes-tendency}
    \frac{\left \langle \d_t \uS \right \rangle}{|f \vL|} \sim \frac{| \buS |}{f T | \buL |} \ll 1 \per
\eeq
We conclude that the Lagrangian-mean hypothesis is consistent since all terms in \eqref{uL-momentum} that explicitly involve surface waves are at least $O(\Ro)$ or smaller.

\section{Ocean general circulation simulations with and without explicit surface wave effects}  \label{sec:wave-model}

We pursue empirical validation of the scaling arguments and conclusions in section~\ref{sec:theory} by describing a novel wave-averaged general circulation model, and comparing simulated surface velocity fields between a realistic, typical ``control'' global ocean simulation and a wave-averaged simulation.
The comparison shows that typical general circulation models --- which do not depend explicitly on the ocean surface wave state --- simulate and output Lagrangian-mean currents.
Both the control and wave-averaged general circulation simulations use models based on the Modular Ocean Model 6 (MOM6) following the Geophysical Fluid Dynamics Laboratory (GFDL)'s OM4 configuration \cite{Adcroft-etal-2019-mom6}.

\subsection{Control general circulation model based on MOM6}

Our control MOM6-based general circulation model (GCM) is called ``Ocean~Model~4'', or OM4.
OM4 is a typical GCM that discretizes and time-integrates the horizontal components of the wave-agnostic, implicitly-averaged Boussinesq momentum equation~\eqref{control-momentum}, with hydrostatic balance
\beq \label{control-hydrostatic-balance}
    \d_z p = b \com
\eeq
approximating the vertical component of \eqref{control-momentum}.

\subsection{A wave-averaged MOM6}

Our wave-averaged GCM, dubbed ``OM4-CL'' \cite<CL after>{craik1976rational} discretizes and time-integrates the horizontal components of the wave-averaged Craik--Leibovich Boussinesq momentum equation~\eqref{uL-momentum}.
OM4-CL replaces the vertical component of equation~\eqref{uL-momentum} with ``wavy hydrostatic balance'' \cite{suzuki2016understanding}
\beq \label{wavy-hydrostatic-balance}
    \d_z \bar p = \bar b \; {\color{PineGreen} - \; (\uL \d_z \uS + \vL \d_z \vS) } \per
\eeq
In OM4-CL, tracers are advected by $\buL$.
Mass conservation is enforced by requiring that $\buL$ is divergence-free, which \citeA{vanneste2022stokes} show is a valid approximation in the CL equations.
OM4-CL follows advances of many previous works on the averaged effects of surface waves on ocean circulation
\cite{mcwilliams1999wave, suzuki2016understanding, couvelard2020development} to implement the wave-averaged equations in an ocean general circulation model.
However, OM4-CL is unique in simulating the Lagrangian-mean velocity directly.
Simulating the Lagrangian-mean velocity is mathematically equivalent to simulating the Eulerian-mean velocity, but simplifies the model implementation due to the structural similarities between \eqref{uL-momentum} and \eqref{control-momentum} (as will be described in detail in a forthcoming documentation paper).

\subsection{Coupled sea ice--ocean model simulations}

Both the control OM4 and the wave-averaged OM4-CL simulations follow the approach for coupled ocean and sea-ice model initialization and forcing laid out by \citeA{Adcroft-etal-2019-mom6}.
Prescribed atmospheric and land forcing fields in these simulations are obtained from the JRA55-do reanalysis product \cite{Tsujino:2018}, following recommendations from the second Ocean Model Intercomparison Project protocol \cite<OMIP2, see>{OMIPa_2016, Tsujino:2020}.
Simulations are performed with a nominal lateral resolution of $\quarterdegree$ that partially resolves mesoscale eddies.
Our configuration is similar to OM4p25 described by \citeA{Adcroft-etal-2019-mom6}, including a hybrid vertical coordinate with two meter vertical resolution near the surface that transitions to isopycnal coordinates in the ocean interior.
Both OM4 and OM4-CL use the same wave-dependent surface boundary layer vertical mixing parameterization \cite{reichl2019parameterization} with the same modeled Stokes drift input.
We conduct simulations using forcing from 1958-2017 and analyze model output from the last 20 years (1998-2017).

For the wave-averaged simulations, global Stokes drift velocities are taken from an offline WAVEWATCH-III v6.07 simulation \cite{WW3DG}, following a similar procedure to \citeA{Li2019,Reichl:2020}.
The Stokes drift profile is reconstructed from five exponentially-decaying surface Stokes drift wavenumber bands \cite<equation (9) in >{Li2019}, which closely matches the profile of Stokes drift obtained from the full wavenumber spectrum (here simulated using 25 discrete wavenumber bands).
The wave model has a horizontal resolution of about 50km, is forced with the same JRA55-do wind product as the ocean, and includes simulated sea-ice provided from the standard OM4 configuration.
For simplicity we neglect the feedback effects of ocean currents on advecting and refracting waves, which are not expected to qualitatively effect our conclusions due to the coarse model scales used in these experiments.

Note that the same winds --- not the same wind stress --- force both OM4 and OM4-CL, and that OM4-CL includes the Stokes tendency term $\d_t \buS$ in equation~\eqref{uL-momentum}.
As a result, OM4 and OM4-CL have slightly different column-integrated momentum budgets \cite{fan2009effect, wagner2021near}.
Nevertheless, figures~\ref{fig:1} and~\ref{fig:2} show that these discrepancies are not important.

\subsection{A conspicuous correlation between instantaneous wave-agnostic and Lagran\-gian-mean currents}

\begin{figure}
    \begin{center}
    \includegraphics[width=1\textwidth]{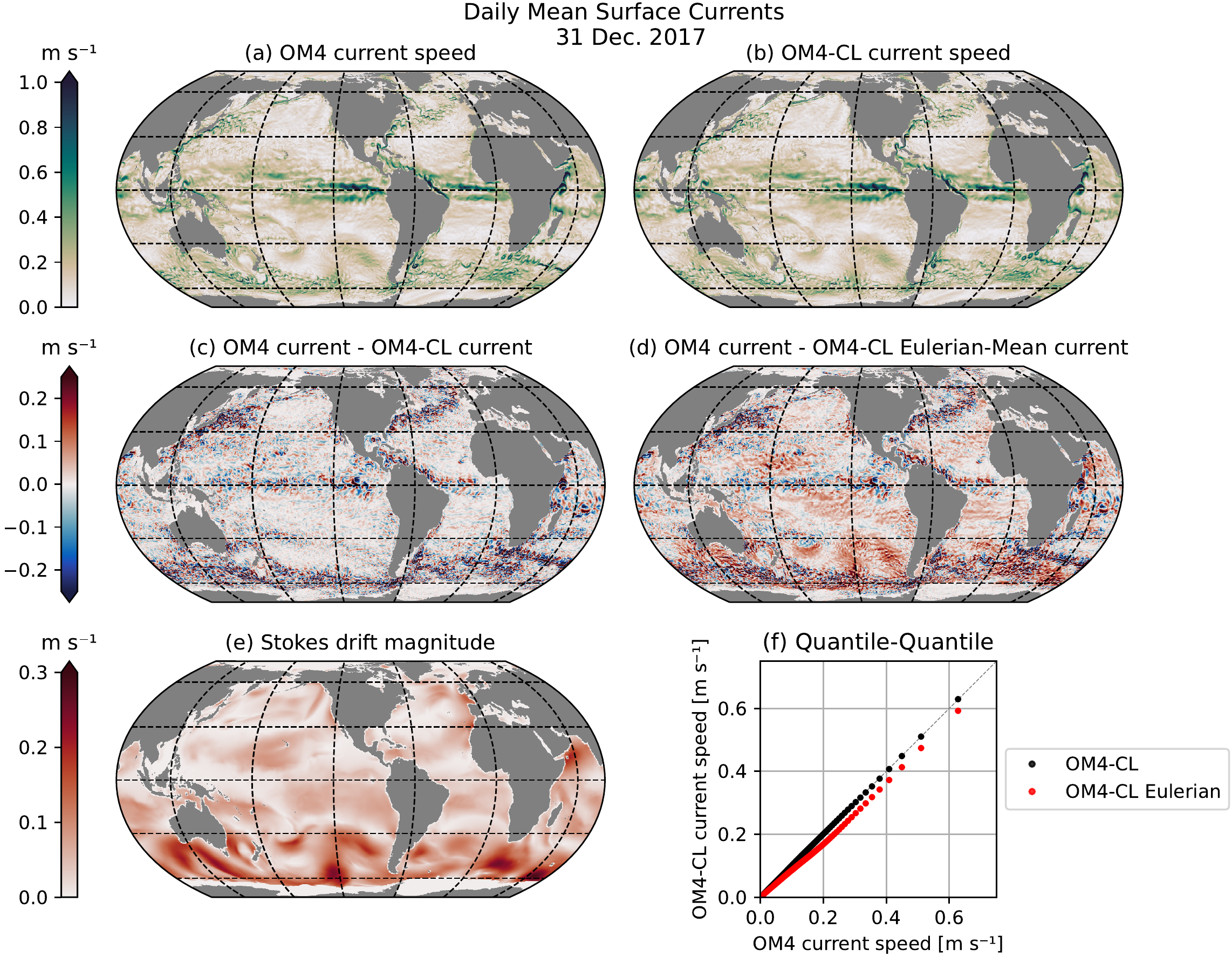}
    \end{center}
    \caption{Comparison of daily-mean snapshot of surface currents in OM4 and OM4-CL. (a)~surface current speed in OM4. (b)~surface current speed in OM4-CL. (c)~Difference of the surface current speed in OM4 (panel (a)) and OM4-CL (panel (b)). (d)~Difference of the surface current speed in OM4 (panel (a)) and the Eulerian mean current from OM4-CL. (e)~Stokes drift current magnitude. (f)~The comparison of the quantile--quantile plot for surface current speeds between \emph{(i)}~OM4-CL and OM4 and \emph{(ii)}~Eulerian mean OM4-CL and OM4. The signature of Stokes drift surface current~(e) is evident in panel~(d).}
    \label{fig:A}
\end{figure}

Figure~\ref{fig:A} compares daily-averaged surface currents in OM4 and OM4CL.
Figure~\ref{fig:A}c shows that the difference between OM4 and OM4-CL surface currents is noisy, which we attribute to chaotic mesoscale variability.
But the difference between OM4 surface currents and \textit{Eulerian-mean} surface currents from OM4-CL, plotted in figure~\ref{fig:A}d, nevertheless exhibits a large-scale signal amidst the mesoscale noise suspiciously similar to the Stokes drift in figure~\ref{fig:A}e.

To quantify this large-scale discrepancy, we plot correlation of the corresponding quantiles in figure~\ref{fig:A}f.
Figure~\ref{fig:A}f demonstrates that the OM4 surface current is perfectly correlated with the Lagrangian-mean current of OM4-CL, while betraying a systematic mismatch between OM4-CL's Eulerian-mean current, validating the Lagrangian-mean hypothesis for OM4.
To eliminate the differences in surface currents due to mesoscale eddy variability, we investigate multi-decadal time-averages of the surface currents in OM4 and OM4-CL.

\subsection{Time-averaged wave-agnostic currents are almost identical to time-averaged Lagrangian-mean currents}

\begin{figure}
    \begin{center}
    \includegraphics[width=1\textwidth]{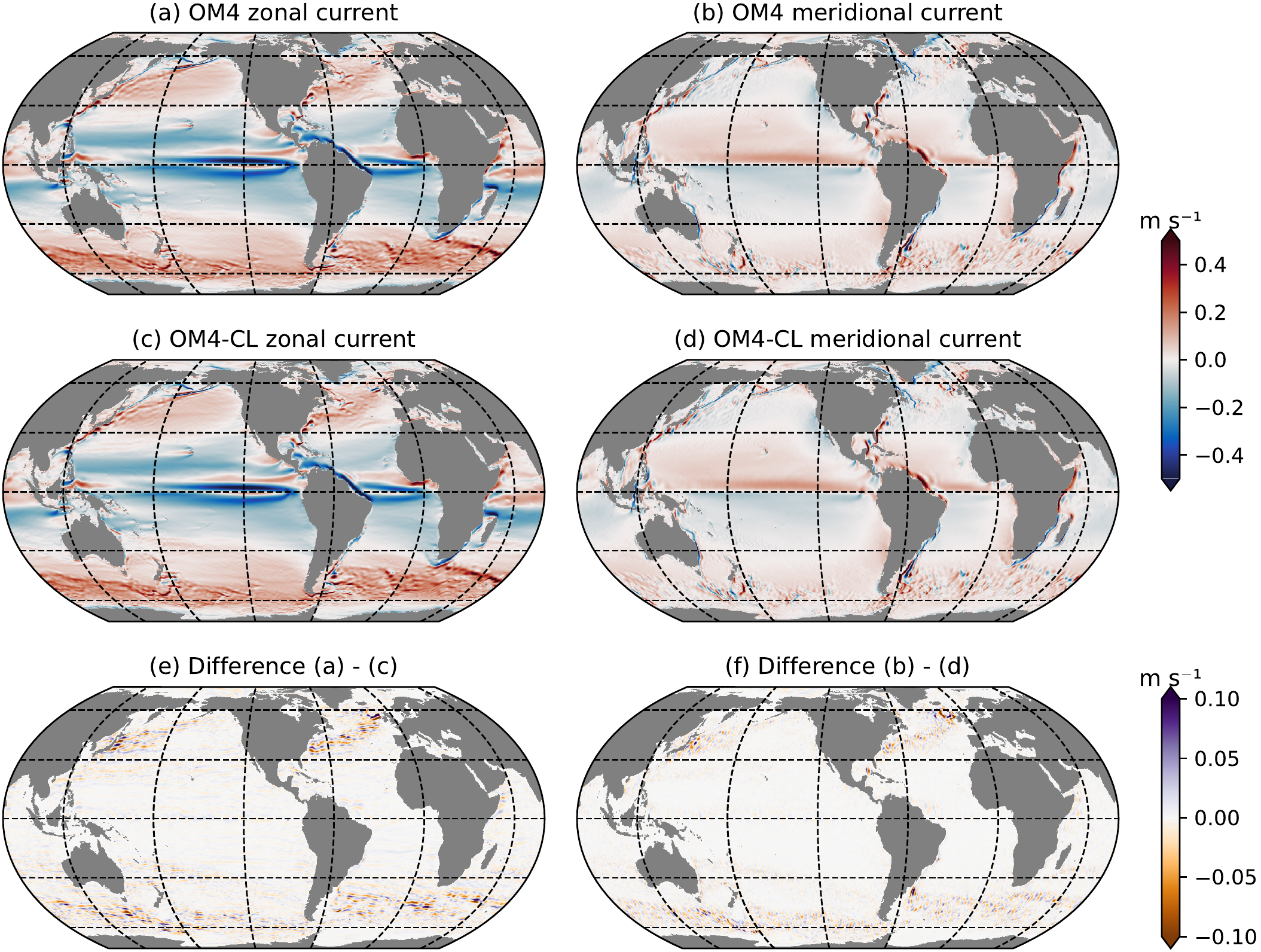}
    \end{center}
    \caption{(Upper) OM4 zonal and meridional surface currents ($\bu$ in \eqref{control-momentum}) averaged between 1998-2017, (middle) time-averaged OM4-CL zonal and meridional currents ($\buL$ in \eqref{uL-momentum}), and (bottom) differences between the upper two rows.
    Note the different color scales between panels (a)-(d) and the bottom two panels.}
    \label{fig:1}
\end{figure}

Figure~\ref{fig:1} compares surface currents between the control OM4 and the wave-averaged OM4-CL.
OM4 simulates ``implicitly-averaged'' currents with no explicit surface wave dependence, while OM4-CL explicitly simulates Lagrangian-mean surface currents.
Currents output from both OM4 and OM4-CL are further averaged over the time period 1998-2017.
The similarity of figure~\ref{fig:1}a-b, which show zonal and meridional components of $\bu$ from OM4, and figure~\ref{fig:1}c-d, which show the zonal and meridional components of the Lagrangian-mean $\buL$ from OM4-CL, demonstrate that the surface circulation in OM4 and the Lagrangian-mean surface circulation in OM4-CL are almost identical.
The differences between the zonal and meridional components of $\bu$ and $\buL$, shown in the bottom row of figure~\ref{fig:1}, are small and associated with chaotic mesoscale perturbations.

\begin{figure}
    \begin{center}
    \includegraphics[width=1\textwidth]{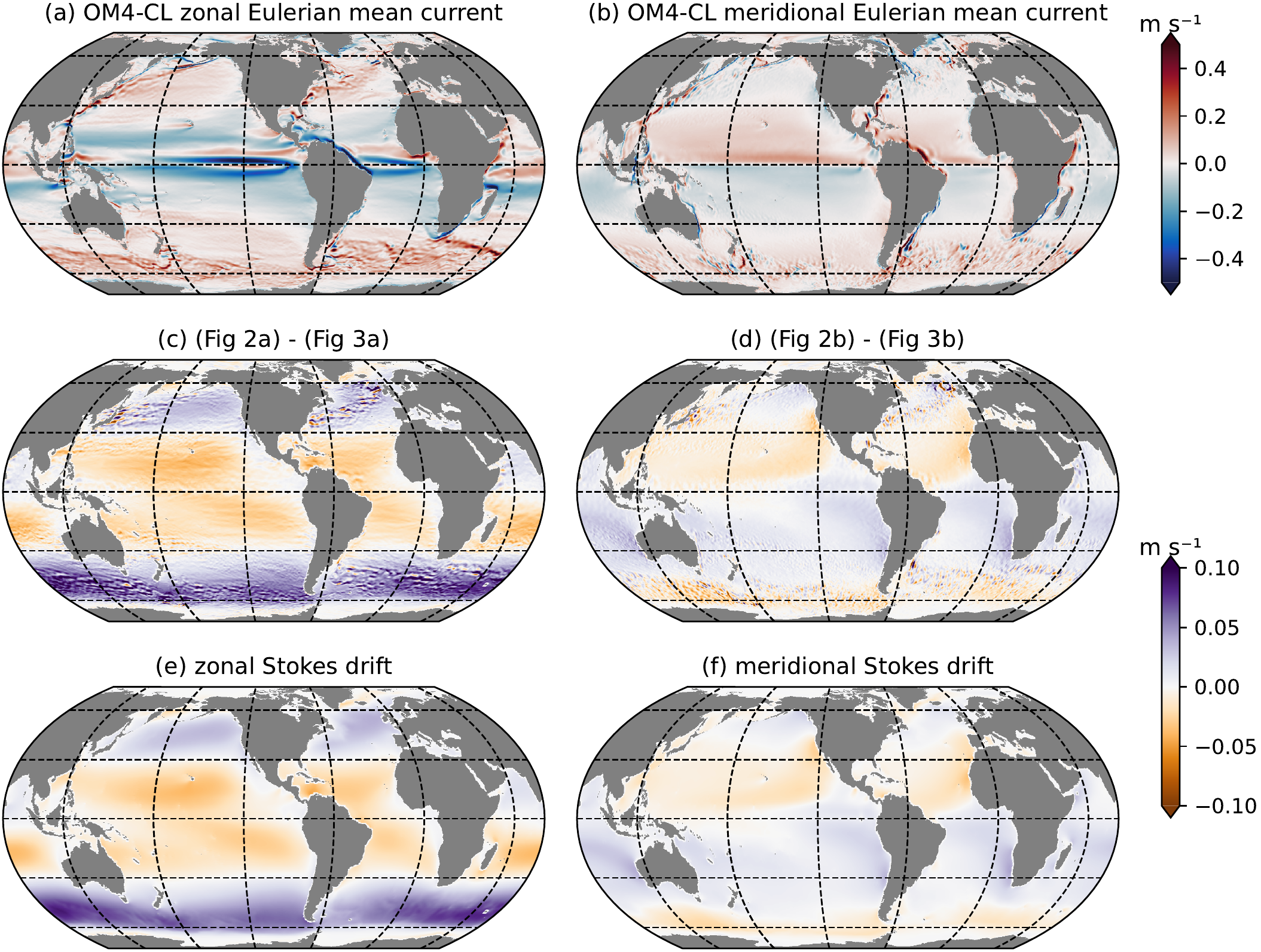}
    \end{center}
    \caption{(Upper) Mean OM4-CL zonal and meridional Eulerian-mean surface currents (1998-2017), (middle) difference between OM4-CL Eulerian mean currents and OM4 currents, and (bottom) mean surface Stokes drift.} \label{fig:2}
\end{figure}

Finally, we directly evaluate the Eulerian-mean hypothesis within OM4 and OM4-CL.
The Eulerian-mean velocity is calculated from OM4-CL output by subtracting the Stokes drift from the simulated velocity $\buL$ according to \eqref{total-velocity}.
The Eulerian-mean hypothesis posits that the mean velocity in the control OM4 simulation is close or identical to Eulerian-mean velocity from the OM4-CL simulation.
However, the middle row of figure~\ref{fig:2} reveals a systematic and significant difference between the Eulerian-mean velocity from OM4-CL and the wave-agnostic velocity from OM4 which is much larger than the differences exhibited in the bottom row of figure~\ref{fig:1}.
Furthermore, the difference between the currents from the control simulation and the Eulerian-mean currents from the wave-averaged simulation (middle row of figure~\ref{fig:2}) turns out to be almost identical to the mean surface Stokes drift currents (bottom row of figure~\ref{fig:2}).
We thus do not find evidence to support the Eulerian-mean hypothesis.
Instead, the current simulated by the wave-agnostic OM4 is close to the Lagrangian-mean current simulated by OM4-CL, as predicted by the Lagrangian-mean hypothesis.

\section{Discussion} \label{sec:discussion}

By inspecting the wave-averaged equations of motion, and comparing the output from wave-neglecting control simulation and an explicitly wave-averaged simulation, we come to three conclusions.
First, the Eulerian-mean hypothesis does not encapsulate a consistent theoretical argument, and therefore Stokes drift should not be added to general circulation model output.
Second, ocean GCMs with horizontal resolutions of $1/4^\circ$ and coarser (and thus with geostrophically-balanced velocities) may accurately simulate the surface-wave-averaged \textit{Lagrangian-mean} velocity.
Finally, resolved --- not parameterized --- surface wave effects are negligible at large oceanic scales where Coriolis forces dominate.

We note that the Eulerian-mean hypothesis assumes the Eulerian-mean velocity field is ``fundamental'', and that the Lagrangian-mean velocity is a diagnostic quantity that can be computed by combining the Eulerian-mean velocity obtained from a dynamical balance and a Stokes drift velocity diagnosed from the surface wave field.
Our results advance a different paradigm: the Lagrangian-mean velocity is the fundamental mean quantity to be determined from the mean momentum balance, while the Eulerian-mean velocity is diagnostic, and ``dependent'' on the surface wave field.
The fundamental nature of the Lagrangian-mean or ``total'' momentum balance is also emphasized by \citeA{samelson2022wind} using averages in a surface-following coordinate system similar to the Lagrangian-mean.

\subsection{Boundary layer parameterization in general circulation models}

General circulation models that solve the Lagrangian-mean equations use --- explicit or implicitly --- parameterizations formulated in terms of $\buL$.
For example, both OM4 \cite{reichl2019parameterization} and the $K$-profile parameterization \cite{large1994oceanic} model the turbulent vertical flux of horizontal momentum with
\beq \label{kpp}
    \b{\X} \approx \d_z \left ( K \, \d_z \buL \right ) \com
\eeq
where the turbulent vertical diffusivity $K$ is a nonlinear function of mean buoyancy $\bar b$, mean velocity $\buL$, surface boundary conditions, and depth $z$.
We emphasize that the parameterization in equation~\eqref{kpp} is sensible, as it dissipates mean kinetic energy $\half | \buL |^2$ \cite{wagner2021near} and is consistent with large eddy simulation results.
For example, \citeA{reichl2016langmuir} find momentum fluxes aligned with $\d_z \buL$ in large eddy simulations of hurricane-forced boundary layer turbulence, and \citeA{pearson2018turbulence} observe that turbulent mixing beneath surface waves tends to homogenize $\buL$.


%

\subsection{Future applications of wave-averaged general circulation models}

Figures~\ref{fig:1} and \ref{fig:2} show that resolved surface wave effects are negligible at $\quarterdegree$ degree resolution.
However, we expect that resolved surface wave effects become more relevant at finer resolutions and higher Rossby numbers, when the term $\left ( \bnabla \times \buS \right ) \times \buL$ in \eqref{uL-momentum} is no longer negligible.
The question remains: ``At what resolution do wave effects matter for mesoscale or submesoscale dynamics?''
Surface wave effects are known to be important at the $O(1 \, \rm{m})$ scales of ocean surface boundary layer large eddy simulations \cite{mcwilliams1997langmuir}, but the effects of surface wave on motions with scales between $O(1 \, \rm{m}$) and $\quarterdegree$ remains relatively unexplored \cite{hypolite2021surface, suzuki2016surface}.

Even $\quarterdegree$-resolution GCMs benefit from knowledge of the surface wave state when their boundary layer turbulence parameterizations depend on the surface wave state \cite{Li2019}.
This is also true for air-sea flux parameterizations \cite{Reichl:2020} and potentially other parameterizations, such as those for wave-ice interaction.


\appendix

\section*{Open Research}

The MOM6 source code including modifications for MOM6-CL is available at \url{https://github.com/mom-ocean/MOM6}.
WAVEWATCH~III source code is available from \url{https://github.com/NOAA-EMC/WW3}.
Code and model output used for generating figures are available at \url{https://github.com/breichl/MOM6CL-Figures} (and will be linked to Zenodo upon acceptance).

\acknowledgments
Without implying endorsement, we gratefully acknowledge discussions and even material assistance from Alistair Adcroft, Brandon Allen, Keaton Burns, Raffaele Ferrari, Glenn Flierl, Baylor Fox-Kemper, Stephen Griffies, Andy Hogg, Adele Morrison, Callum Shakespeare, William Young, and many others --- not least of all the late Sean Haney.
Be careful, Sean.
G.L.W.~is supported by the generosity of Eric and Wendy Schmidt by recommendation of the Schmidt Futures program, and by the National Science Foundation under grant~AGS-6939393.
N.C.C.~is supported by the Australian Research Council DECRA Fellowship DE210100749.
We extend additional thanks to Zoom Video Communications for allowing collaboration during the difficult times of the COVID-19 pandemic.

\end{document}